# Properties of graphene deposited on GaN nanowires: influence of nanowire roughness, self-induced nanogating and defects

Jakub Kierdaszuk,[1,*] Piotr Kaźmierczak,[1] Justyna Grzonka,[2] Aleksandra Krajewska,[2,3] Aleksandra Przewłoka,[2] Wawrzyniec Kaszub,[2] Zbigniew R. Zytkiewicz,[4] Marta Sobanska,[4] Maria Kamińska,[1] Andrzej Wysmołek,[1] Aneta Drabińska,[1]

[1]Faculty of Physics, University of Warsaw, ul. Pasteura 5, 02-093, Warsaw, Poland

[2]Institute of Electronic Materials Technology, ul. Wólczyńska 133, 01-919, Warsaw, Poland

[3]Institute of Optoelectronics, Military University of Technology, ul. Gen. Sylwestra Kaliskiego 2, 01-476, Warsaw, Poland

[4]Institute of Physics, Polish Academy of Sciences, Al. Lotnikow 32/46, 02-668, Warsaw, Poland

______________________________

[*]To whom correspondence should be addressed. E-mail: jakub.kierdaszuk@fuw.edu.pl

ABSTRACT

We present detailed Raman studies of graphene deposited on gallium nitride nanowires with different variations in height. Our results show that different density and height of nanowires being in contact with graphene impact graphene properties like roughness, strain and carrier concentration as well as density and type of induced defects. Detailed analysis of Raman spectra of graphene deposited on different nanowire substrates shows that bigger differences in nanowires height increase graphene strain, while higher number of nanowires in contact with graphene locally reduce the strain. Moreover, the value of graphene carrier concentration is found to be correlated with the density of nanowires in contact with graphene. Analysis of intensity ratios of Raman G, D and D' bands enable to trace how nanowire substrate impacts the defect concentration and type. The lowest concentration of defects is observed for graphene deposited on nanowires of the lowest density. Contact between graphene and densely arranged nanowires leads to a large density of vacancies. On the other hand, grain boundaries are the main type of defects in graphene on rarely distributed nanowires. Our results also show modification of graphene carrier concentration and strain by different types of defects present in graphene.

INTRODUCTION

Combination of excellent electrical and mechanical properties with interesting physical phenomena occurring in two-dimensional structure makes graphene an interesting experimental material to study.[1,2,3] Importantly, it is a promising material for new kind of low dimensional transistors, gas sensors, ultra-capacitors, electrodes for solar cells, and for van der Waals heterostructures. In order to construct these devices an interaction between graphene and adjacent layers should be recognized. It is well established already that graphene grown on silicon carbide is less strained on substrate terraces than on terrace edges, while electron concentration on the edges is lower than on terraces.[4] This example shows that fluctuations of substrate morphology substantially modify graphene properties.

Novel approach of graphene-based nanostructures are van der Waals heterostructures in which graphene is transferred onto another material with different morphology and electronic properties.[5] However in that kind of structures several aspects, like for example strain induced by mechanical contact between materials or gating of graphene by neighbouring layers, are important for further applications. Furthermore, electron scattering on defects modifies graphene properties in several ways, for example, additional scattering centres reduce carrier mobility and consequently graphene conductivity. On the other hand, chemical functionalization of graphene may improve the sensitivity of graphene-based sensors.[6] Therefore control of density and types of defects in graphene might be a new way to prepare efficient molecular sensors.

Systems containing graphene on nanowires have been used in solar cells to increase their efficiency. In particular, it has been shown that application of nanowires in solar cells decreases the light reflection by scattering of light in-between nanowires.[7,8] Nanowires have also high cross-section of light absorption.[9] However, the interaction between corrugated nanowire substrate and graphene could substantially increase the scattering of carriers in graphene electrode and decrease its conductivity. Therefore, detailed studies of the interaction between nanowire substrate and graphene are crucial to gain a deep understanding of phenomena occurring on that interface. One of the most common experimental techniques for studying properties of graphene is Raman spectroscopy.[10] Non-invasive measurements of inelastic light scattering give an insight into phonon structure of graphene. Analysis of graphene G and 2D band parameters provides an information about a number of graphene layers, strain and carrier concentration.[11,12,13,14,15] Furthermore, in defected graphene D and D' defect bands are also observed with intensities related to concentration of defects and their types. [16,17,18,19,20] Thus, careful studies of Raman spectra allow to determine how substrate impacts graphene properties and consequently modifies efficiency of graphene-based structures.

In this paper, we present detailed studies of Raman spectra of graphene deposited on gallium nitride nanowires (GaN NWs) with different variations in height. Electric field induced in GaN predicted by theoretical calculations could reach 5 MV/cm.[21] This is an effect of high spontaneous and piezoelectric polarizations in the wurtzite structure of GaN. Consequently, high concentration of carriers on the GaN surface can be observed.[22,23]

Previous studies of graphene on GaN NWs have shown that electric charges located on the top of the GaN NWs strongly impact Raman scattering in graphene, causing enhancement of the spectrum.[24,25] Therefore, studies of graphene on NWs with different density and variations in height might give an information about the role of the density of supporting points under graphene on its properties. For example, analysis of graphene deposited on uniformly distributed silicon nanopillars showed the dependence of graphene strain on distances between nanopillars.[26] For small distances graphene was clearly suspended, while graphene ripples caused by strain in samples with larger distances between pillars were observed. Nevertheless, nanowire substrate could also gate graphene and effect carrier concentration and its distribution in the layer. Coulomb interaction between GaN NWs and graphene could also create the vacancies in graphene and consequently increase the density of defects. In turn, as reported recently, strain and carrier concentration can be influenced and modified by graphene defects as well.[27,28,29,30] Therefore, determination of how nanowire morphology, nanogating and Coulomb interaction impact graphene properties is important, not only for basic research but also for future applications of such structures. This requires determination of the influence of graphene interaction with the NWs substrate on graphene properties what is the main topic of this work.

EXPERIMENTAL DETAILS

Monolayer graphene was grown by Chemical Vapour Deposition (CVD) technique on a copper foil with methane gas as a precursor.[31] Next graphene was transferred onto GaN NWs substrates. Due to low adhesive force between graphene and corrugated substrate the most common method of transfer graphene with use of PMMA (poly(methyl methacrylate)) polymer could not be applied for the transfer onto NWs substrate.[31] Therefore we used stable orthogonal frame from PDMS (polydimethylsiloxane) polymer to stabilize graphene during the transfer process.[32] GaN NWs substrates were fabricated by Plasma Assisted Molecular Beam Epitaxy (PAMBE) method in N rich conditions on (111) silicon substrate.[33] Application of different growth temperatures and growth times allowed to obtain nanowires with different variations in height.[34] In our experiment, we used three samples differing in NWs substrate height and density. Their detailed parameters are included in Table I. In the first sample (thereafter named as N0) NWs had similar height of ~900 nm and their average density was ~140 $\mu m^{-2}$. However, they formed clusters containing several merged NWs. In the second sample (N100) the height of NWs varied by ~100 nm, from 300 to 400 nm, and the density of NWs was about 400 NWs $\mu m^{-2}$. In the third sample (N500) the average density of NWs was similar to that in the first sample – about 120 NWs $\mu m^{-2}$. However, in this sample, two distinct groups of NWs were observed – about 80% of them were 1 $\mu$m in height, while about 20% reached 1.5 $\mu$m.

The samples were studied by scanning electron microscopy (SEM) using SU8230 Hitachi microscope equipped with an in-lens secondary electron detector at 5 kV electron beam voltage. The Raman spectra were collected by T64000 Horiba Jobin-Yvon spectrometer with Nd:YAG laser operating at 532 nm wavelength as the excitation source, and objective with a magnification of 100× that allowed to obtain the spatial resolution of approximately 300 nm. The micro-Raman maps were collected with 100 nm step with a few square micrometres of mapping area for each sample. The spectra were calibrated by a reference sample of high quality silicon.

RESULTS AND DISCUSSION

The morphology of graphene deposited on NWs with different variations in height is presented in Figure 1. The large cracks visible in graphene layer are caused by the transfer process. Graphene on NWs with equal height is smooth (Fig. 1a, d). Small wrinkles are the evidence of the small expansion of graphene hanging in-between individual NWs. Larger wrinkles are observed in graphene on NWs with 100 nm variations in height (Fig. 1b, e). Nevertheless, due to the higher density of supporting points graphene is still attached to every single nanowire including those slightly lower in height. The most expanded graphene is observed in the N500 sample (Fig. 1c, f). Opposite to the others, in this case, graphene touches only the highest NWs and does not have any contact with the lowest ones. Furthermore, graphene in N100 and N500 samples is pierced by some of the highest NWs. The topography of graphene on NWs with different variations in height is also visualized in schematic profiles (Fig. 1g-i). SEM result suggested that both parameters, namely differences in NWs' height and density of NWs under graphene affect graphene morphology. Therefore, according to our previous results a higher number of NWs in contact with graphene may increase an effect of nanogating while a lower number of supporting points could increase graphene strain.[26,35]

Analysis of representative Raman spectra for each sample shows that both graphene bands (G band at about 1585 cm$^{-1}$ and 2D band at 2680 cm$^{-1}$) and both defects bands (D band at 1345 cm$^{-1}$ and D' band at 1620 cm$^{-1}$) are observed (Figure 2). In order to recognize how NWs locally modify graphene strain, carrier concentration and defects, a statistical analysis of band parameters over the whole Raman micro-mapping area was performed. In our system the lowest mapping step is comparable with the average distance between NWs and three times smaller than the diameter of a laser spot. Therefore, single measurement is averaged over a few NWs and local interactions between graphene and small groups of NWs are traced rather than interactions with a single NW. Graphene strain can be studied by analysis of the position of 2D band energy and its full-width at half maximum (FWHM). Dependence of graphene strain on 2D band energy shift has been described by the equation: [36]

$$E_{2D} = E_{2D}^0 - 2\gamma_{2D} E_{2D}^0 \Delta\varepsilon, \qquad (1)$$

where $\gamma_{2D}$ is the Grüneisen parameter, $\Delta\varepsilon$ is a value of strain, and the value of 2D band energy for unstrained graphene ($E_{2D}^0$) has been reported to be 2677.6 cm$^{-1}$.[14] Positive value of $\Delta\varepsilon$ corresponds to tensile strain while negative to the compressive one. Grüneisen parameter determines the rate of change of a given phonon frequency in a crystal in respect to strain. Its value depends on the strain type and substantial differences between values of the Grüneisen parameter for uniaxial and biaxial strain were observed.[14,37,38,39,40] Thus a description of strain in the structure of graphene deposited on a large number of supporting points is not straightforward. Consequently, we cannot calculate the absolute value of strain, however its qualitative description is still possible. The 2D band has a complex line-shape due to the double resonance signal.[41,42] Therefore, graphene strain could be qualitatively discussed by analysis of 2D band FWHM.[14,43]

Histograms of 2D band energy are presented in Figure 3a, e, i, while calculated average values of 2D band energy and their standard deviations are presented in Table II. Interestingly, for graphene transferred onto NWs with equal height (N0 sample), the strain has tensile character (Fig 3a), while in graphene on NWs with different variations in height (N100 and N500 samples) strain is rather compressive (Fig. 3e, i). The highest value of average 2D band energy (2690.2 cm$^{-1}$) is observed for N500 sample, while the highest standard deviation of 2D band energy (2.7 cm$^{-1}$) is observed for N100 sample (Tab. II). Therefore, we can conclude that the highest strain in N500 sample is related to the extension of graphene between rarely arranged supporting points while the biggest local strain fluctuations are observed for graphene transferred on densely arranged NWs with medium differences in height (N100 sample). We suppose that strain in graphene on NWs is uniaxial or biaxial only in a local scale, between nearest NWs. However, in a Raman experiment, the excitation beam of 300 nm diameter probes the larger area containing several NWs. Due to the random distribution of NWs, total character of strain is neither simple biaxial nor uniaxial. Therefore, we cannot calculate the absolute value of graphene strain, but estimation of its value and comparison between samples is still possible. Table II presents the calculated average values of strain for all samples using on the Grüneisen parameter equal to 0.012 as obtained by Mohiuddin et al.[14] Interestingly, the average absolute value of strain for N0 and N100 sample is the same and equals to 0.07%. It is three times lower than for N500 sample where strain reaches 0.2%. These results are further confirmed by analysis of 2D band FWHM (Fig 3b, f, j). The average value of 2D FWHM for N0 and N100 samples is comparable, although slightly lower for N0 sample. On the other hand, for N500 sample 2D FWHM is significantly higher. This result confirmed the presence of high strain in N500 sample. According to the analysis of 2D energy, different values of 2D band FWHM for N0 and N100 samples cannot be explained by the effect of graphene strain only, but has to be caused by other reasons like, for example carrier mobility. 2D band energy and FWHM also depends on carrier concentration. However, their changes are significantly lower than found for the G band FWHM what will be discussed in the next paragraph.[44]

Analysis of graphene G band parameters allows to trace how NW substrate impacts carrier concentration. G band energy dependence on carrier concentration is described by the equation:

$$E_G = E_G^0 - 2\gamma_G E_G^0 \Delta\varepsilon + n \cdot 7.38 \cdot 10^{13} \qquad (2)$$

where $\gamma_G$ is the Grüneisen parameter for G band and *n* is carrier concentration in cm$^{-2}$.[36] $E_G^0$ is a value of G band energy for unstrained and undoped graphene which was found to be equal to 1583.5 cm$^{-1}$.[14] Sensitivity of G band energy on carrier concentration is caused by the presence of Kohn anomaly near the Γ point in phonon band structure of graphene.[15] Consequently, G band energy significantly increases with increasing both electron or hole concentration.[45] Unfortunately, G band energy depends not only on carrier concentration but also on the strain, therefore for estimation of the value of carrier concentration in strained graphene analysis of the values of both G and 2D band parameters is necessary. Another parameter which depends on graphene carrier concentration is the FWHM of G band.[15] Phonon lifetime is short in case of low value of Fermi energy. Thus, band width following uncertainty principle consequently becomes larger. Increasing the Fermi energy leads to increase of the phonon lifetime and consequently to decrease of band width. In general, FWHM of G band is positively correlated with the value of graphene strain, but in case of graphene with strain smaller than 0.2%, which is the case in our samples, such changes of FWHM are negligible.[46]

Histograms of G band energy and its FWHM are presented in Figure 3. The average value of G band energy for N0 and N500 is the same and equals to 1584.4 cm$^{-1}$ (Tab. II), while for N100 sample it is 4.2 cm$^{-1}$ higher. A similar trend can be observed in the standard deviation of G band energy. For N100 sample it is significantly higher than for N0 and N500 samples. On the other hand, average G band FWHM is similar for N0 and N100 samples and significantly lower than observed for N500 sample. Interestingly, the standard deviation of G band FWHM for N100 sample is more than four times higher than for N0 and N500 samples. As it was discussed before, the existence of medium tensile strain should decrease the value of G band energy in N0 sample. Similarly, compressive strain observed in N100 sample should increase the value of G band energy. Analysis of characteristic values of Grüneisen parameters for different types of strain shows that strain-induced change of G band energy is less than twice smaller than change observed for the 2D band. However, values of G band energy in N0 and N100 samples are about 3 cm$^{-1}$ higher than expected from the strain impact. Considering low value of G band FWHM for both samples, changes of G band energy in N0 and N100 sample could be explained by the higher carrier concentration in these samples than in N500. The lowest value of G band FWHM is present in N100 sample which suggested the highest carrier concentration among all investigated samples.

Two factors should be taken into account when explaining our results. Firstly, differences in NWs height and their density impact graphene elongation and consequently affect graphene strain. Higher differences in NWs height in N500 sample increase graphene strain while larger density of GaN/graphene supporting points in N100 sample is responsible for the local reduction of strain. Secondly, GaN nanowire substrate modifies graphene carrier concentration by self-induced nanogating.[35] Local carrier concentration in graphene on NWs is higher than in graphene between NWs. A large number of NWs in contact with graphene in N0 and N100 samples increases the value of carrier concentration. Our results suggest also that low density of NWs contacting graphene in N500 sample is responsible for the low value of carrier concentration. Therefore, the density of NWs supporting graphene could be responsible for the observed values of strain and carrier concentration. Moreover, high values of standard deviation for G and 2D band energies and FWHM for N100 sample is probably caused by local fluctuations of NWs height in densely arranged NWs. Therefore, strain and carrier concentration in N100 sample change significantly between data points. Moreover, a higher value of 2D band FWHM in N100 sample suggests different carrier mobilities in N0 and N100 samples (Fig. 3b, f).

The intensity ratio of 2D and G graphene bands in monolayer graphene has been reported to be negatively correlated with carrier concentration.[15] Higher Fermi energy increases the probability of scattering on free carriers, which adds to scattering on phonons. Consequently, the intensity ratio of 2D and G Raman bands, $R_{2DG}$, decreases when carrier concentration increases. Histograms of $R_{2DG}$ for all measured samples are presented in Figure 4. The highest standard deviation is observed for N100 sample, which is 1.5 times higher than for N0 sample and 6 times higher than for N500 sample (see Table II). Average value of the $R_{2DG}$ ratio is the highest for N0 sample (5.2), and the lowest for N500 (1.7). What is surprising, the value of $R_{2DG}$ suggests that the carrier concentration in N500 sample is the highest from all investigated samples which disagree with the conclusions obtained from analysis of the G and 2D band energies and FWHM. In order to clarify that contradiction, analysis of dependency of $R_{2DG}$ on strain and the carrier concentration is performed by analysis of 2D and G band FWHM. As it was discussed before, 2D band FWHM is positively correlated with graphene strain, while G band FWHM is negatively correlated with carrier concentration.[14,15] Negative correlation between $R_{2DG}$ ratio and 2D band FWHM is observed (Fig. 5a). Therefore, it can be concluded that $R_{2DG}$ decreases when graphene strain increases. On the other hand, analysis of $R_{2DG}$ dependence on G band FWHM does not show any evident correlation (Fig. 5b). Experimental points for each sample are separated from each other. Thus, our results suggest that the intensity ratio of 2D and G band in graphene on NWs is correlated rather with strain than with carrier concentration, which is in contradiction to results reported by Das et al.[15]

Other kinds of graphene Raman bands visible in spectra in Fig. 2 are D and D' bands – so-called defects bands. In the case of graphene transferred onto NWs analysis of scattering on defects allows to trace how graphene structure changes after deposition on NWs and how these changes depend on the density of NWs and their differences in height. An additional aspect is defect impact on graphene strain and carrier concentration. Experimental studies have shown that some kinds of defects distort graphene lattice and consequently increase graphene strain.[47,48] For example vacancies elongate graphene lattice and induce tensile strain, while Stone-Wales defects reduce the bonds length which results in compressive strain in graphene. On the other hand, a large number of vacancies may relax strain in expanded graphene.[27] Additionally, a disorder in graphene influences its carrier concentration. In case of the low density of defects, an increase of disorder is correlated with increasing carrier concentration and the sign of charge carriers depends on defect type.[49,30,29] For example vacancies and nitrogen dopants in nitrilic and pyridinic position introduce p-type doping while nitrogen dopants in graphitic position and hydrogen dopants in pyridinic result in n-type doping.[50] Thus defect origin and density impact graphene strain, carrier concentration as well as interaction with the substrate.

G band is generated by scattering on iTO or iLO phonon near Γ point of the Brillouin zone. For the presence of D band, resonant scattering on iTO phonon near K point of the Brillouin zone and the defect is necessary. Consequently, the intensity of G band is proportional to the area of laser spot while the intensity of D band depends on the number of defects in the excited area. Therefore density of defects ($n_D$) is inversely proportional to an intensity ratio of G and D bands ($R_{GD}$) and described by the equation: [19]

$$n_D(\mu m^{-2}) = 1.8 \cdot 10^{14} [\lambda_l(nm)]^{-4} R_{GD}^{-1}, \qquad (3)$$

where $\lambda_l$ is a wavelength of excitation light. In order to visualize the distribution of defect density on graphene surface, we performed spatial and statistical analysis of intensity ratios of G and D bands (Fig. 6). The respective 2D maps of $R_{GD}$ ratio are presented in Figures 6a-c. Distribution of $R_{GD}$ ratio in graphene on NWs with equal height is rather plain while in N100 is slightly modulated by interaction with the NW substrate. More evident modulation of $R_{GD}$ parameter is observed in N500 sample. Figures 6 d-f show histograms of $R_{GD}$ ratio while the average value of $R_{GD}$ and density of defects calculated using equation 3 are presented in Table III. Analysis of histograms presented in Figure 6d and e shows that average value of $R_{GD}$ and width of distribution are comparable in the N0 and N100 samples (Tab. III). The average density of defect distribution in N0 and N100 sample is about 977, and 936 defects per square micrometre, respectively, while the density of nanowires under graphene in N0 sample is three times lower than in N100. Average value of $R_{GD}$ in N500 sample is twice higher (Tab. III). Consequently, the average density of defects is twice lower than in the previous two samples and is equal to 449 per square micrometre. This observation suggested that very low density of supporting points (24 NWs per $\mu m^2$) is correlated with the low density of defects.

However, a similar number of defects in N0 and N100 remains unclear. Although for N0 and N100 samples density of defects is similar, different distribution of $R_{GD}$ ratio which reflects defect density on the surface is observed (fig. 6a-c). In case of graphene transferred onto NWs with equal height, clusters of NWs locally interact with graphene stronger than areas between them, whereas large density of NWs with medium variations in height introduces a modulation of defect density in N100 sample. Moreover, a low number of supporting points in N500 sample is correlated with the lower average density of defects. However, the $R_{GD}$ ratio in N500 sample is densely modulated on the mapping area and does not reflect the supporting NW pattern. This result suggests that deformation of graphene hanging between rarely distributed NWs also creates defects which explains $R_{GD}$ behaviour shown on Figure 6c. Therefore, our results suggest that not only contact between NWs and graphene but also graphene deformation itself create defects in graphene and influence their spatial distribution. Very low density of supporting NWs also decreases the number of defects in graphene.

The intensity of both defect bands D and D' ($R_{DD'}$) depends on defect density and parameters describing perturbation introduced by defects in the crystal lattice. These perturbation parameters depend on the type of defect and are different for D and D' bands. Thus intensity ratio between D and D' bands characterize the type of defects in graphene.[20] Previous experimental results have shown that the value of $R_{DD'}$ ratio equal to 3.5 is characteristic for grain boundaries, 5 for multiple vacancies, 7 corresponds to single vacancies, while 13 is observed for $sp^3$ hybridisation defects.[20,51] Furthermore, theoretical calculations predicted values of 1.3 and 10.5 for on-site and hopping defects, respectively.[52] In order to identify the types of defects present in the studied samples, the intensity ratio between D and D' bands is analysed (Fig. 7). In contrast to the $R_{GD}$ ratio, strong modulation of $R_{DD'}$ by NWs substrate is observed on 2D maps (Fig 7a-c) for all the samples. This observation suggests that nanowire substrate directly impacts the observed types of defects.

Histograms of $R_{DD'}$ ratio are presented in Figure 7d-f. Gauss distributions corresponding to the types of defects were fitted for each histogram. Percentage contribution of defects of a specific type for each sample are calculated by dividing the area of each Gauss distribution by the sum of areas of all fitted Gauss distributions. Type of defects and their percentage contribution are included in Table III. Interestingly, for all samples, one maximum of high intensity and several smaller maxima can be observed, and about 80% or more defects are described by the main maximum. Single vacancies are dominant defects in N0 and N100 samples (maximum of distribution at $R_{DD'}$ equal to 8.3 and 7.5 respectively) while grain boundaries are the main defects in N500 sample (maximum of distribution at $R_{DD'}$ equal to 4.1). At least 98% of all types of defects in N500 sample are grain boundaries, which is a higher value than obtained for vacancy contribution in N0 and N100 samples (88% and 79%, respectively). The standard deviation of $R_{DD'}$ ratio for the main maximum in N500 sample is equal to 0.6, which is lower than for N0 and N100 sample (0.8 and 1.3 respectively). Therefore, the interaction between graphene and rarely distributed NWs is more homogenous than with densely arranged NWs. The largest number of different types of defects – five – are observed in N100 sample which confirms that graphene interacts with densely distributed NWs in a variety of ways.

As discussed before, analysis of 2D band energy and $R_{GD}$ ratio shows that N0 and N100 samples are characterized by a similar average absolute value of strain and similar density of defects. In the case of N500 sample, for which strain is significantly higher, a higher value of $R_{GD}$ and lower value of $R_{DD'}$ ratio are found. In Fig. 8 we present dependence of $R_{DD'}$ on $R_{GD}$ mapped points in Raman experiment in all studied samples. The negative correlation of $R_{DD'}$ and $R_{GD}$ is observed. Therefore, the dependency of $R_{DD'}$ and $R_{GD}$ ratios on G and 2D band energy and FWHM were detailed studied in order to trace interdependence between disorder parameters and carrier concentration or strain. No explicit correlations between $R_{GD}$ and $R_{DD'}$ ratios and carrier concentration were found. However, both parameters are correlated with 2D band FWHM and consequently with graphene strain for all investigated samples. Local stretching of graphene observed in investigated samples should rather elongate graphene lattice than create new defects. Therefore, the lower density of NWs supporting graphene in N500 sample is responsible for the lower density of defects and higher strain. The higher density of defects in N0 and N100 sample is caused by the higher density of NWs under graphene, however, the reason of a different kind of strain (tensile/compressive) in these two samples is unclear. From the discussion above, the dependence between the density of NWs supporting graphene and types of defects is nontrivial. Our results suggest, that in graphene deposited on rarely arranged NWs grain boundaries are the most dominant type of defects. Densely arranged nanowire substrate introduce vacancies in graphene deposited on them. Furthermore, the presence of a large number of vacancies in N0 and N100 samples together with gating by nanowire substrate could be responsible for increasing of carrier concentration, which is confirmed also by other studies.[50,35] In N500 sample where most of defects are grain boundaries less number of bonds were cracked. Consequently, a number of carriers is lower than in samples with a significant presence of vacancies. Vacancies could also increase local tensile strain in graphene similarly as the nanowire substrate.

Therefore, high differences in NWs height and low density of supporting points decrease the observed density of defects and highlight grain boundaries defects omnipresent in graphene layers. Contact with NWs of lower differences in height and higher density of supporting points creates more vacancies and increases their density on the surface. Moreover, graphene strain and carrier concentration could be locally modified by the different density of defects and their types. Thus, further studies on influence of NWs supporting graphene and graphene strain, carrier concentration and defects performed with higher resolution are essential.

CONCLUSIONS

We transferred graphene on GaN NWs with 0, 100 and 500 nm variations in height, and studied their properties by SEM and Raman spectroscopy. Graphene on NWs with different variations in height is rippled and pierced by the highest NWs. Detailed analysis of Raman spectra shows that differences in NWs height as well as their density strongly impact graphene strain and carrier concentration. The highest strain is observed for sample with the highest, 500 nm, differences in height. Unexpectedly strain in graphene on NWs with equal height has tensile while in graphene on NWs with non-equal height compressive character. Analysis of G band energy and G band FWHM shows a positive correlation between the density of NWs under graphene and value of carrier concentration. In contradiction to previous reports, we found that intensity ratio between 2D and G band is correlated rather with graphene strain than with carrier concentration. Furthermore, analysis of $R_{GD}$ and $R_{DD'}$ ratios showed that the density of defects in graphene was affected by nanowire substrate. Our results suggested that NWs supporting graphene with low differences in height introduce vacancies in graphene. Increasing distances between NWs decrease the density of defects and expose a larger number of grain boundaries omnipresent in any graphene. Furthermore, vacancies could locally increase graphene carrier concentration and tensile strain in N0 and N100 samples together with nanowire substrate. Thus, the density of NWs supporting graphene substrate and their differences in height impact graphene carrier concentration and strain. It is also possible to think about the use of NW substrate for defect engineering.

ACKNOWLEDGEMENTS

ACKNOWLEDGEMENTS

This work was partially supported by the Ministry of Science and Higher Education in years 2015-2019 as a research grant "Diamond Grant" (No. DI2014 015744). GaN nanowires were grown within the Polish National Science Centre Grants No. UMO-2016/21/N/ST3/03381 and 2016/23/B/ST7/03745. This work was supported by the Research Foundation Flanders (FWO) under Grant n° EOS 30467715.

BIBLIOGRAPHY

[1] A.K. Geim and K.S. Novoselov, Nat. Mater. **6**, 183 (2007).

[2] A.K. Geim, Science **324**, 1530 (2009).

[3] K.S. Novoselov, V.I. Fal′ko, L. Colombo, P.R. Gellert, M.G. Schwab, and K. Kim, Nature **490**, 192 (2012).

[4] K. Grodecki, R. Bozek, W. Strupinski, A. Wysmolek, R. Stepniewski, and J.M. Baranowski, Appl. Phys. Lett. **100**, 261604 (2012).

[5] A.K. Geim and I. V. Grigorieva, Nature **499**, 419 (2013).

[6] K.-J. Huang, D.-J. Niu, J.-Y. Sun, C.-H. Han, Z.-W. Wu, Y.-L. Li, and X.-Q. Xiong, Colloids Surfaces B Biointerfaces **82**, 543 (2011).

[7] G. Fan, H. Zhu, K. Wang, J. Wei, X. Li, Q. Shu, N. Guo, and D. Wu, ACS Appl. Mater. Interfaces **3**, 721 (2011).

[8] H. Park, S. Chang, J. Jean, J.J. Cheng, P.T. Araujo, M. Wang, M.G. Bawendi, M.S. Dresselhaus, V. Bulović, J. Kong, and S. Gradečak, Nano Lett. **13**, 233 (2013).

[9] P. Krogstrup, H.I. Jørgensen, M. Heiss, O. Demichel, J. V Holm, M. Aagesen, J. Nygard, and A. Fontcuberta i Morral, Nat. Photonics **7**, 306310 (2013).

[10] A.C. Ferrari, J.C. Meyer, V. Scardaci, C. Casiraghi, M. Lazzeri, F. Mauri, S. Piscanec, D. Jiang, K.S. Novoselov, S. Roth, and A.K. Geim, Phys. Rev. Lett. **97**, 187401 (2006).

[11] L.G. Cançado, A. Reina, J. Kong, and M.S. Dresselhaus, Phys. Rev. B **77**, 245408 (2008).

[12] L.M. Malard, J. Nilsson, D.C. Elias, J.C. Brant, F. Plentz, E.S. Alves, A.H. Castro Neto, and M.A. Pimenta, Phys. Rev. B **76**, 201401 (2007).

[13] J. Zabel, R.R. Nair, A. Ott, T. Georgiou, A.K. Geim, K.S. Novoselov, and C. Casiraghi, Nano Lett. **12**, 617 (2012).

[14] T.M.G. Mohiuddin, A. Lombardo, R.R. Nair, A. Bonetti, G. Savini, R. Jalil, N. Bonini, D.M. Basko, C. Galiotis, N. Marzari, K.S. Novoselov, A.K. Geim, and A.C. Ferrari, Phys. Rev. B - Condens. Matter Mater. Phys. **79**, 205433 (2009).

[15] A. Das, S. Pisana, B. Chakraborty, S. Piscanec, S.K. Saha, U. V Waghmare, K.S. Novoselov, H.R. Krishnamurthy, A.K. Geim, A.C. Ferrari, and A.K. Sood, Nat. Nanotechnol. **3**, 210 (2008).

[16] R. VIDANO and D.B. FISCHBACH, J. Am. Ceram. Soc. **61**, 13 (1978).

[17] R.J. Nemanich and S.A. Solin, Phys. Rev. B **20**, 392 (1979).

[18] F. Tuinstra and L. Koenig, J. Chem. Phys. **53**, 1126 (1970).

[19] L.G. Cançado, A. Jorio, E.H.M. Ferreira, F. Stavale, C.A. Achete, R.B. Capaz, M.V.O. Moutinho, A. Lombardo, T.S. Kulmala, and A.C. Ferrari, Nano Lett. **11**, 3190 (2011).

[20] A. Eckmann, A. Felten, A. Mishchenko, L. Britnell, R. Krupke, K.S. Novoselov, and C. Casiraghi, Nano Lett. **12**, 3925 (2012).

[21] O. Ambacher, J. Smart, J.R. Shealy, N.G. Weimann, K. Chu, M. Murphy, W.J. Schaff, L.F. Eastman, R. Dimitrov, L. Wittmer, M. Stutzmann, W. Rieger, and J. Hilsenbeck, J. Appl. Phys. **85**, 3222 (1999).

[22] T. Takeuchi, C. Wetzel, S. Yamaguchi, H. Sakai, H. Amano, I. Akasaki, Y. Kaneko, S. Nakagawa, Y. Yamaoka, and N. Yamada, Appl. Phys. Lett. **73**, 1691 (1998).

[23] N. Jamond, P. Chrétien, F. Houzé, L. Lu, L. Largeau, O. Maugain, L. Travers, J.C. Harmand, F. Glas, E. Lefeuvre, M. Tchernycheva, and N. Gogneau, Nanotechnology **27**, 325403 (2016).

[24] J. Kierdaszuk, P. Kaźmierczak, A. Drabińska, K. Korona, A. Wołoś, M. Kamińska, A. Wysmołek, I. Pasternak, A. Krajewska, K. Pakuła, and Z.R. Zytkiewicz, Phys. Rev. B **92**, 195403 (2015).

[25] J. Kierdaszuk, P. Kaźmierczak, R. Bożek, J. Grzonka, A. Krajewska, Z.R. Zytkiewicz, M. Sobanska, K. Klosek, A. Wołoś, M. Kamińska, A. Wysmołek, and A. Drabińska, (2017).

[26] A. Reserbat-Plantey, D. Kalita, L. Ferlazzo, S. Autier-Laurent, K. Komatsu, C. Li, R. Weil, Z. Han, A. Ralko, L. Marty, S. Guéron, N. Bendiab, H. Bouchiat, and V. Bouchiat, Nano Lett. **14**, 5044 (2014).

[27] G. López-Polín, C. Gómez-Navarro, V. Parente, F. Guinea, M.I. Katsnelson, F. Pérez-Murano, and J. Gómez-Herrero, Nat. Phys. **11**, 26 (2014).

[28] A. Das, B. Chakraborty, and A.K. Sood, Bull. Mater. Sci. **31**, 579 (2008).

[29] M. Hofmann, Y.-P. Hsieh, K.-W. Chang, H.-G. Tsai, and T.-T. Chen, Sci. Rep. **5**, 17393 (2015).

[30] A. Ferrari and J. Robertson, Phys. Rev. B **61**, 14095 (2000).

[31] T. Ciuk, I. Pasternak, A. Krajewska, J. Sobieski, P. Caban, J. Szmidt, and W. Strupinski, J. Phys. Chem. C **117**, 20833 (2013).

[32] I. Pasternak, a. Krajewska, K. Grodecki, I. Jozwik-Biala, K. Sobczak, and W. Strupinski, AIP Adv. **4**, 097133 (2014).

[33] A. Wierzbicka, Z.R. Zytkiewicz, S. Kret, J. Borysiuk, P. Dluzewski, M. Sobanska, K.

Klosek, A. Reszka, G. Tchutchulashvili, A. Cabaj, and E. Lusakowska, Nanotechnology **24**, 035703 (2013).

[34] M. Sobanska, K.P. Korona, Z.R. Zytkiewicz, K. Klosek, and G. Tchutchulashvili, J. Appl. Phys. **118**, 184303 (2015).

[35] J. Kierdaszuk, P. Kaźmierczak, R. Bożek, J. Grzonka, A. Krajewska, Z.R. Zytkiewicz, M. Sobanska, K. Klosek, A. Wołoś, M. Kamińska, A. Wysmołek, and A. Drabińska, Carbon N. Y. **128**, 70 (2018).

[36] J.M. Urban, P. Dąbrowski, J. Binder, M. Kopciuszyński, A. Wysmołek, Z. Klusek, M. Jałochowski, W. Strupiński, and J.M. Baranowski, J. Appl. Phys. **115**, 233504 (2014).

[37] I. Calizo, A.A. Balandin, W. Bao, F. Miao, and C.N. Lau, Nano Lett. **7**, 2645 (2007).

[38] C. Metzger, S. Rémi, M. Liu, S. V. Kusminskiy, A.H. Castro Neto, A.K. Swan, and B.B. Goldberg, Nano Lett. **10**, 6 (2010).

[39] N. Mounet and N. Marzari, Phys. Rev. B **71**, 205214 (2005).

[40] N. Ferralis, J. Mater. Sci. **45**, 5135 (2010).

[41] O. Frank, M. Mohr, J. Maultzsch, C. Thomsen, I. Riaz, R. Jalil, K.S. Novoselov, G. Tsoukleri, J. Parthenios, K. Papagelis, L. Kavan, and C. Galiotis, ACS Nano **5**, 2231 (2011).

[42] M. Huang, H. Yan, T.F. Heinz, and J. Hone, Nano Lett. **10**, 4074 (2010).

[43] Y. Zhao, X. Liu, D.Y. Lei, and Y. Chai, Nanoscale **6**, 1311 (2014).

[44] Y. Li, in *Probing Response Two-Dimensional Cryst. by Opt. Spectrosc.* (Springer, Cham, 2016), pp. 9–18.

[45] H.P. Boehm, A. Clauss, G.O. Fischer, and U. Hofmann, Zeitschrift Für Anorg. Und Allg.

Chemie **316**, 119 (1962).

[46] O. Frank, G. Tsoukleri, J. Parthenios, K. Papagelis, I. Riaz, R. Jalil, K.S. Novoselov, and C. Galiotis, ACS Nano **4**, 3131 (2010).

[47] S. Ebrahimi, Solid State Commun. **220**, 17 (2015).

[48] N. Blanc, F. Jean, A. V. Krasheninnikov, G. Renaud, and J. Coraux, Phys. Rev. Lett. **111**, 1 (2013).

[49] A. Das, B. Chakraborty, and A.K. Sood, Bull. Mater. Sci. **31**, 579 (2008).

[50] T. Schiros, D. Nordlund, L. Pálová, D. Prezzi, L. Zhao, K.S. Kim, U. Wurstbauer, C. Gutiérrez, D. Delongchamp, C. Jaye, D. Fischer, H. Ogasawara, L.G.M. Pettersson, D.R. Reichman, P. Kim, M.S. Hybertsen, and A.N. Pasupathy, Nano Lett. **12**, 4025 (2012).

[51] S. Hang, Z. Moktadir, and H. Mizuta, Carbon N. Y. **72**, 233 (2014).

[52] P. Venezuela, M. Lazzeri, and F. Mauri, Phys. Rev. B - Condens. Matter Mater. Phys. **84**, 035433 (2011).

List of Figures

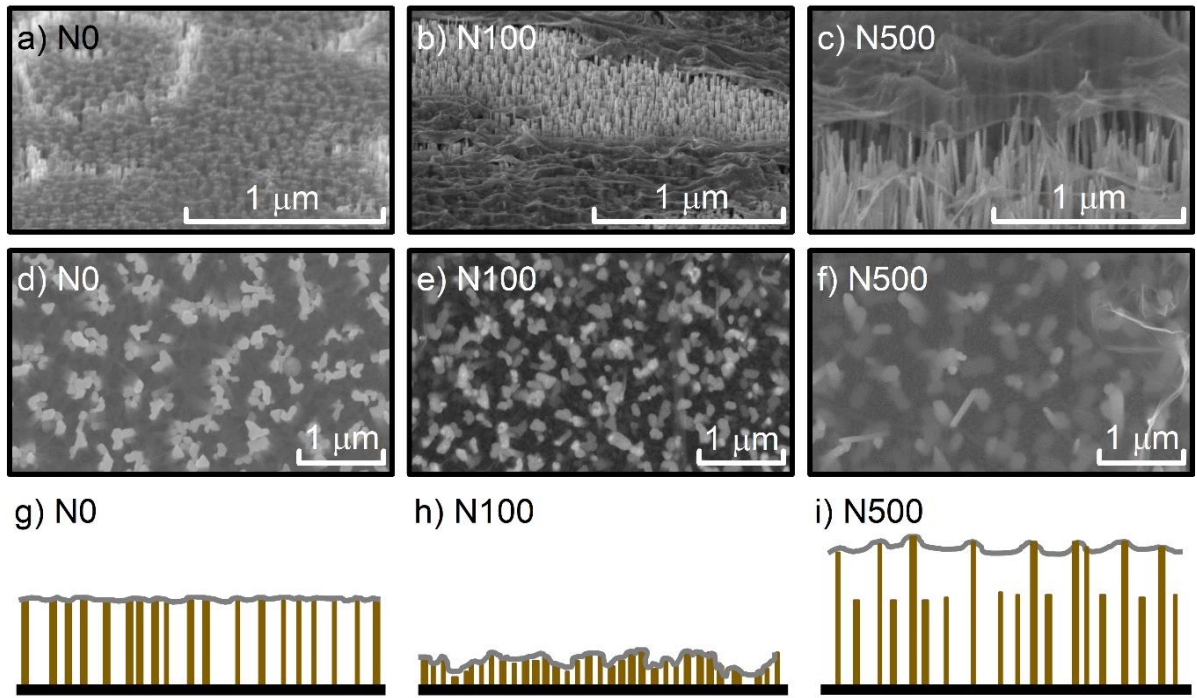

Figure 1. SEM images of graphene on GaN NWs with different variations in height in N0 (a, d), N100 (b, e), and N500 (c, f) samples. (a-c) images were obtained at 70° tilt of the sample while (d-f) were collected in the plan view. Schematic profiles of investigated samples are shown in (g-i).

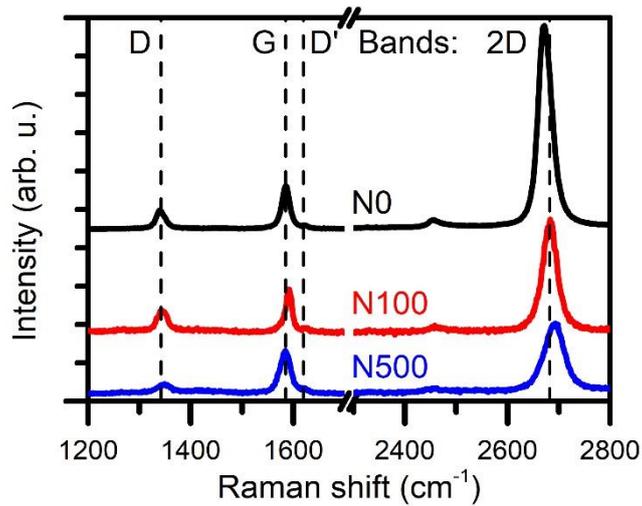

Figure 2. Representative Raman spectra of graphene on NWs with different variations in height normalized to the G band intensity.

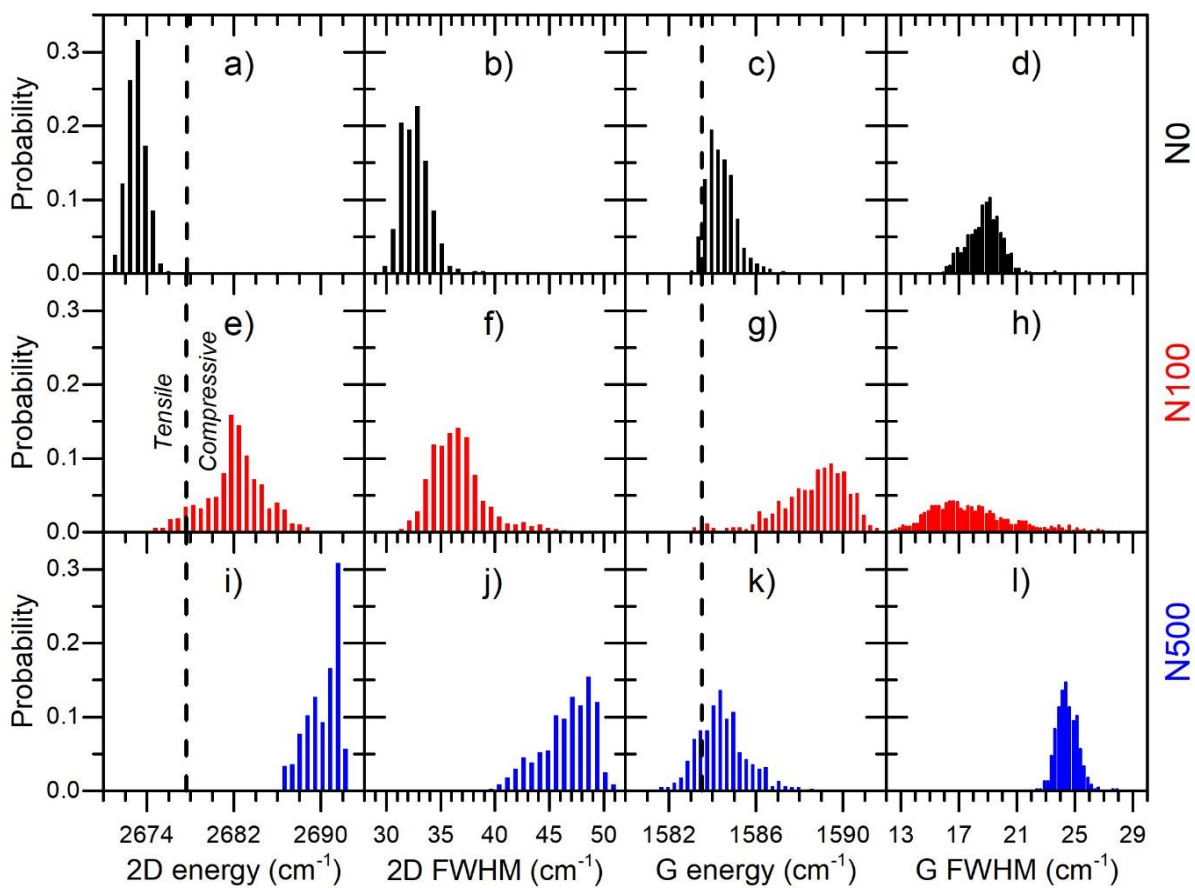

Figure 3. Histograms of 2D band energy (a, e, i), 2D FWHM (b, f, j), G band energy (c, g, k) and G FWHM (d, h, l) for N0, N100 and N500 samples, respectively. Dashed vertical lines

correspond to literature values of 2D and G band energy for unstrained and undoped graphene.[14]

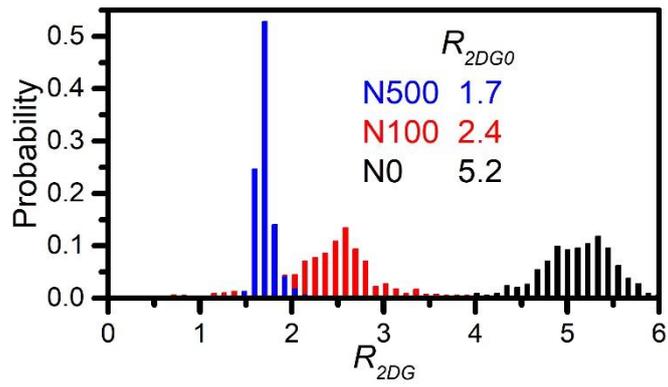

Figure 4. Histograms of intensity ratios between 2D and G bands.

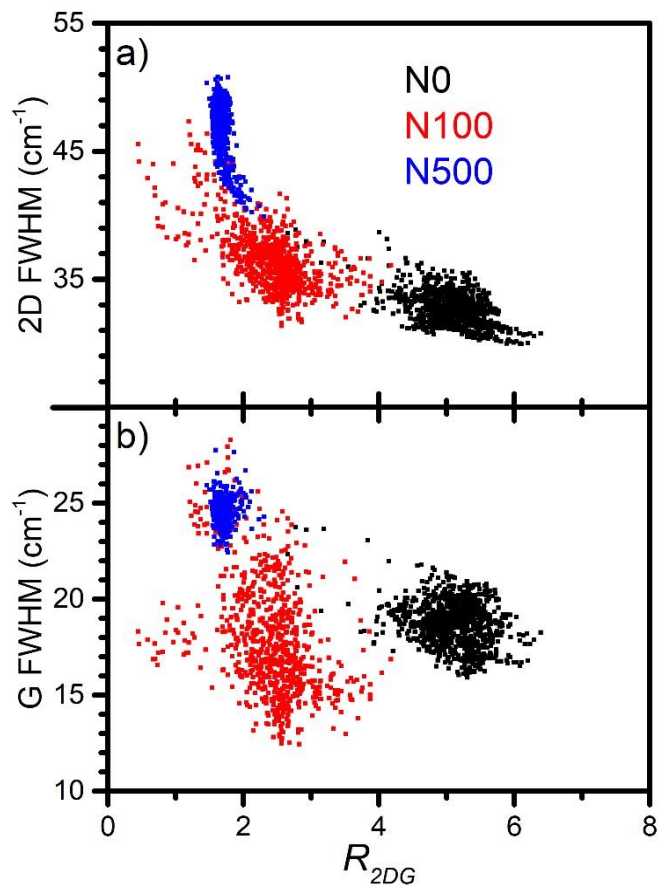

Figure 5. 2D FWHM (a) and G FWHM (b) dependence on the ratio of 2D and G bands intensities ($R_{2DG}$) for graphene on NWs with different variations in height.

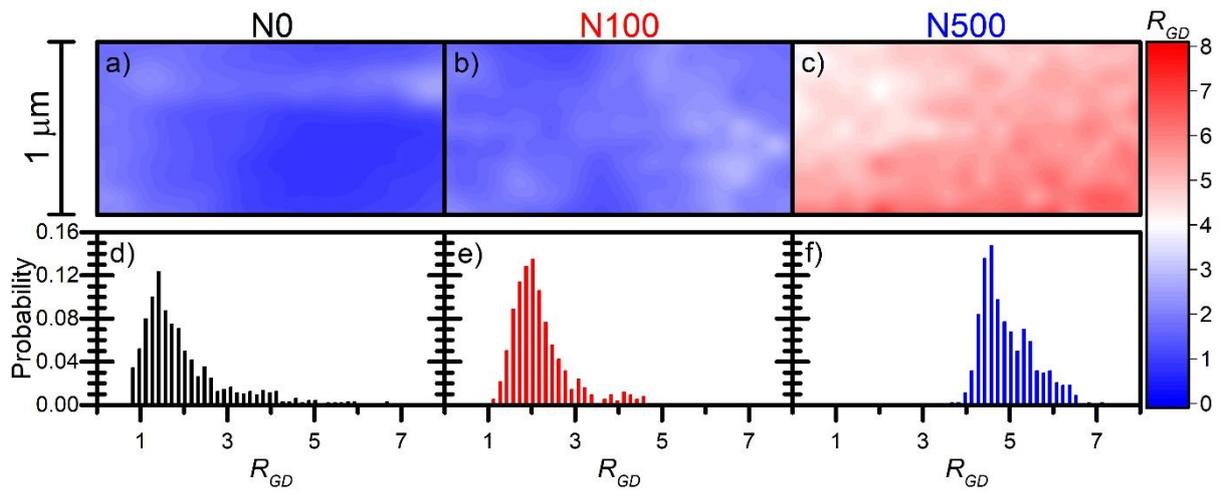

Figure 6. 2D maps (a, b, c) and histograms (d, e, f) of the intensity ratio of G and D bands ($R_{GD}$) for all three samples.

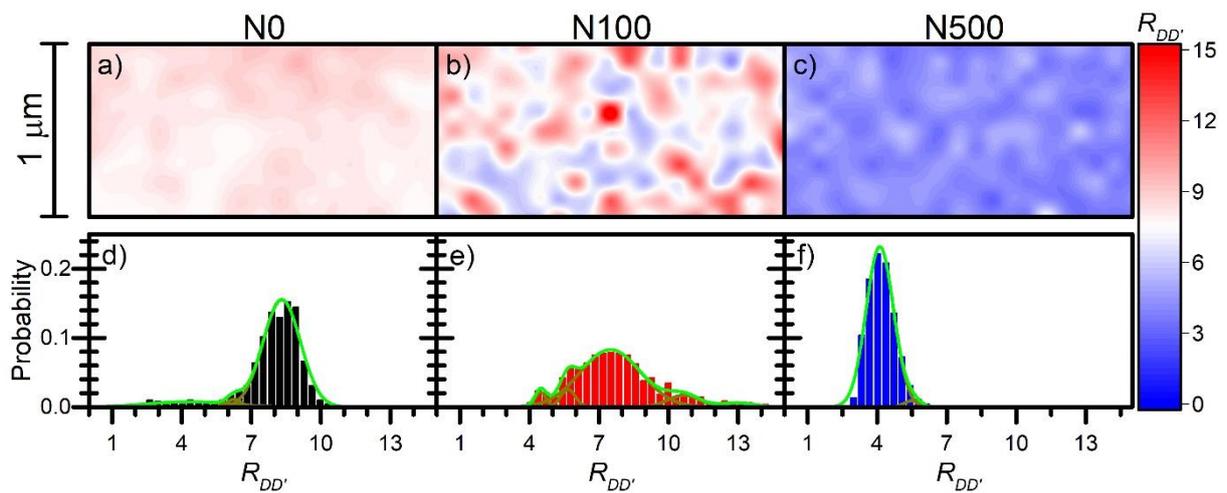

Figure 7. 2D maps (a, b, c) and histograms of the intensity ratio between D and D' band ($R_{DD'}$) for all three samples.

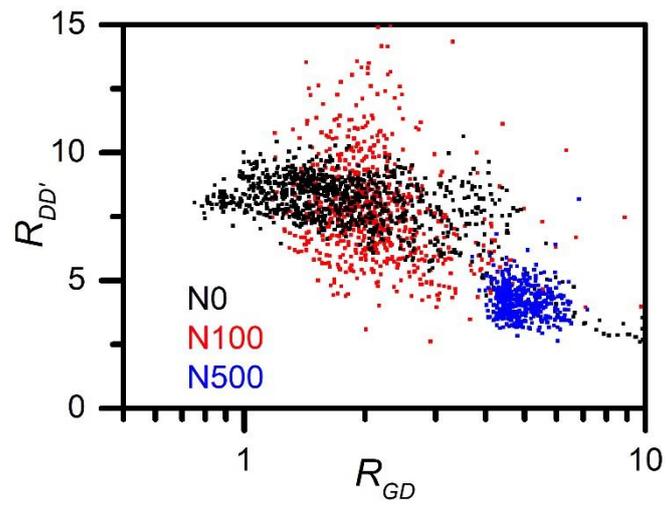

Figure 8. $R_{DD'}$ ratio dependence on $R_{GD}$ for graphene on NWs with different variations in height.

TABLES

Table I, Parameters of three investigated GaN NWs substrates.

| NWs | N0 | N100 | N500 |
|---|---|---|---|
| diameter (nm) | 40 | 40 | 40 |
| height (nm) | 900 | 300-400 | 1000-1500 |
| density of individual NWs ($\mu m^{-2}$) | 140 | 400 | 120 |
| distances between individual NWs (nm) | 80 | 50 | 90 |
| density of NWs clusters ($\mu m^{-2}$) | 20 | 50 | 15 |
| distances between clusters (nm) | 250 | 150 | 260 |

Table II, Average 2D and G band energies ($<E_{2D}>$ and $<E_G>$) with their standard deviation ($\sigma E_{2D}$, $\sigma E_G$), 2D and G band FWHM ($<F_{2D}>$ and $<F_G>$) with their standard deviation ($\sigma F_{2D}$, $\sigma F_G$) and calculated value of average strain ($\Delta\varepsilon$) in graphene on NWs with different variations in height. Positive or negative value of graphene strain corresponds to tensile and compressive strain, respectively.

| | N0 | N100 | N500 |
|---|---|---|---|
| $<E_{2D}>$ (cm$^{-1}$) | 2673.0 | 2682.1 | 2690.2 |
| $\sigma E_{2D}$ (cm$^{-1}$) | 0.9 | 2.7 | 1.5 |
| $\Delta\varepsilon$ (%) | +0.07 | -0.07 | -0.20 |
| $<F_{2D}>$ (cm$^{-1}$) | 32.3 | 36.6 | 46.6 |
| $\sigma F_{2D}$ (cm$^{-1}$) | 1.3 | 2.5 | 2.3 |
| $<E_G>$ (cm$^{-1}$) | 1584.4 | 1588.6 | 1584.4 |
| $\sigma E_G$ (cm$^{-1}$) | 1.2 | 1.6 | 1.1 |
| $<F_G>$ (cm$^{-1}$) | 18.8 | 17.9 | 24.5 |
| $\sigma F_G$ (cm$^{-1}$) | 0.7 | 3.0 | 0.7 |

2929

Table III, Average ratio of G and D bands ($<R_{GD}>$), defect density ($n_D$) and percentage defect identification in graphene on NWs with different variations in height. The most common defect in each sample was highlighted in bold.

|  | N0 | N100 | N500 |
|---|---|---|---|
| $<R_{GD}>$ | 2.3 | 2.4 | 5 |
| $n_D\ (\mu m^{-2})$ | 977 | 936 | 449 |
| grain boundaries | 10% | 4% | **98%** |
| mixture of vacancies | 2% | 6% | 2% |
| single vacancies | **88%** | **79%** | - |
| hopping defects | - | 8% | - |
| sp$^3$ defects | - | 3% | - |